\documentclass[12pt,pre,aps,superscriptaddress]{revtex4}
\usepackage{graphicx}
\usepackage{bm}
\usepackage{amssymb}
\newcommand{\beq}[2]{\begin{equation}#1\label{#2}\end{equation}}
\newcommand{\ceq}[1]{(\ref{#1})}

\newfont{\mbld}{cmbx10 scaled 800}
\newfont{\cab}{cmsy10 scaled 1200}
\newfont{\scab}{cmsy10 scaled 1000}
\newfont{\bcall}{cmbsy10 scaled 1200}

\begin{document}
\title{On a possible approach to general field theories with
  nonpolynomial interactions} 
\author{Franco Ferrari}
\email{ferrari@fermi.fiz.univ.szczecin.pl}
\affiliation{Institute of Physics and CASA*, University of Szczecin,
  ul. Wielkopolska 15, 70-451 Szczecin, Poland}
\begin{abstract}
In this work a class of field theories with self-interactions
described by a potential of the kind $V(\phi(x)-\phi(x_0))$ is
studied. $\phi$ is a massive scalar field and $x,x_0$ are points in a
$d$ dimensional space. Under the condition that the potential admits
the Fourier representation, it is shown that such theories may be
mapped into a standard field theory, in which the interaction of the
new fields is a polynomial of fourth degree.
With some restrictions, this mapping allows the perturbative treatment
of models that are otherwise intractable with standard field theoretical
methods.

A nonperturbative  approach to these theories is attempted. 
The original scalar field $\phi$ is integrated out exactly at the
price of introducing auxiliary vector fields. The latter are treated
in a mean field theory approximation. The singularities that arise
after the elimination of the auxiliary fields
 are cured using the dimensional regularization.
The expression of the counterterms to be subtracted
 is computed.
\end{abstract}
\maketitle
\section{Introduction}\label{sec:intro}
In this letter we study a wide class of $d-$dimensional field
theories in which the interactions are described by a general
potential $V(\phi(x)-\phi(x_0))$. Here $\phi(x)$, $x\in\mathbb{R}^d$
denotes a massive scalar field. $x_0$ is a fixed point in $\mathbb{R}^d$.
The only requirement on the potential $V$ is that its Fourier
representation exists, i.~e. it is possible to write
 $V(\phi)=\int_{-\infty}^{+\infty}db\tilde
V(b)e^{-ib\phi}$.
It is shown that all theories of this kind can be mapped into a
$(d+2)-$dimensional field theory, in which the interactions between
the fields are polynomial. As a consequence, models which are highly
nonlinear and nonlocal may be treated after the mapping
using perturbative methods.
This is the main result of this work.
The mapping is obtained extending a technique known in statistical
mechanics as Gaussian
integration \cite{gi1,gi2,gi2,gi4,gi5}, which allows to identify
certain field theories with a 
gas of interacting particles. In the present case a field
theory is 
identified with another field theory. Yet, Gaussian integration
is used at some step in order to simplify 
the interaction term of the original massive scalar fields.
More precisely, the term $e^{-V}$ is
rewritten in the form
of the ``equilibrium limit'' of the partition
function of a system of quantum
particles interacting with the field $\phi$. A similar strategy has been
recently applied in \cite{FePa} to reformulate the Liouville field
theory as a theory 
with polynomial interactions, which is very similar to scalar
electrodynamics.  A brief introduction to the method and a discussion
of its advantages
can be found in Ref.~\cite{FePa2}.
As a result of the whole procedure, we obtain a theory of complex
scalar fields $\psi^*,\psi$ describing the fluctuations of particles
immersed in the purely longitudinal vector potential $\mathbf
A=\nabla\phi$. 

In the second part of this letter, a nonperturbative approach is
attempted. First of all, the 
field $\phi$ is integrated out using a technique similar to that
exploited in the case of Chern-Simons fields in
Ref.~\cite{Femultifield}. In this way a set of new vector fields
$\bm\xi^*,\bm\xi$ is introduced, which are treated using
a mean 
field theory approximation. The arising singularities are computed
with the help of the dimensional regularization.
This does not exhaust all possible divergences that may arise in the
theory. A discussion of renormalization issues is presented in the
Conclusions. 
\section{The Mapping}\label{sec:gracan}
We consider here the class of $d-$dimensional field theories with
partition function
\beq{Z=\int{\cal D}\phi e^{-S}}{partfun}
and action
\beq{
S=\int d^dx\left[
\frac 12
(\nabla\phi)^2+\frac {m^2}2\phi^2+V(\phi(x)-\phi(x_0))
\right]
}{action}
The potential $V(\phi(x)-\phi(x_0))$ is given in the Fourier
representation:
\beq{V(\phi(x)-\phi(x_0))=
  \int_{-\infty}^{+\infty}db\tilde{V}(b)e^{-ib(\phi(x)-\phi(x_0))}}{vecpotfouref}
One can obtain in this way a wide class of potentials.
For example, the potential:
\beq{
V_1(\phi(x)-\phi(x_0))=\frac {k_1}{a^2+(\phi(x)-\phi(x_0))^2}
}{exaone}
corresponds to the choice
$\tilde{V}(b)=\frac {k_1}{2a}e^{-a|b|}$ with $a>0$.

Putting instead $\tilde{V}_2(b)=\frac{k_2}{\sqrt{4a\pi}}\sin\left(
\frac{b^2}{4a}+\frac{\pi}{4}
\right)$ we have
\beq{
V_2(\phi(x)-\phi(x_0))=k_2\sin\left(
a(\phi(x)-\phi(x_0))^2
\right)
}{exatwo}
Potentials of this kind, which contain in general infinite powers of
the fields as
Eqs.~\ceq{exaone} and \ceq{exatwo} show, can be
simplified with the help of the following identity:
\beq{
\exp\left[
\int_{-\infty}^{+\infty}db \tilde{V}(b)e^{-ib(\phi(x)-\phi(x_0))}
\right]=\lim_{T\to+\infty}\Xi_T[\phi]
}{idefun}
where
\beq{
\Xi_T[\phi]=\int{\cal D}\psi^*{\cal D}\psi\exp\left\{
-i\int dbdtd^dx\left[
i\psi^*\frac{\partial\psi}{\partial
  t}-g|(\nabla+ib\nabla\phi)\psi|^2-J^*\psi-J\psi^*
\right]
\right\}
}{xitidef}
while $\psi=\psi(t,x,b)$ and $\psi^*=\psi^*(t,x,b)$.
In Eq.~\ceq{xitidef} the currents $J^*,J$ have been chosen as follows:
\beq{
J^*(t)=(4\pi ig T)^{\frac d2}\delta(t)\qquad\qquad
J(t,x,b)=-\delta(x-x_0)\tilde{V}(b)\delta(t-T)
}{curcho}
Let us prove the above identity. The complex field $\psi^*$ in
Eq.~\ceq{xitidef} is a Lagrange multiplier that imposes the condition:
\beq{
i\frac{\partial \psi}{\partial t}+g(\nabla+ib\nabla\phi)^2\psi=J
}{condlagrmult}
The solution of this equation is:
\beq{
\psi(t,x,b)=\int db'dt'd^dx'{\cal G}(t-t',x-x',b-b')J(t',x',b')
}{solucondlagrmult}
where
\beq{
{\cal G}(t-t',x-x',b-b')=-\frac{i\theta(t-t')}{|4\pi ig(t-t')|^{\frac d2}}
\exp\left[i\frac{(x-x')^2}{4g(t-t')}
\right]e^{-ib\phi(x)}e^{ib'\phi(x')}\delta{(b-b')}
}{greefunc}
and $\theta(t-t')$ is the Heaviside function. As a consequence, it is
not difficult after integrating out the fields $\psi^*$ and $\psi$ to
show that
\beq{
\Xi_T[\phi]=\exp\left\{
+\int db
dtd^dxdt'dx'J^*(t,x){\cal G}_0(t-t',x-x')e^{-ib\phi(x)}e^{ib\phi(x')}J(t,x,b) 
\right\}
}{xitipsifin} 
In the above equation we have put for convenience:
\beq{
{\cal G}_0(t-t',x-x')=-\frac{i\theta(t-t')}{|4\pi ig(t-t')|^{\frac d2}}
\exp\left[
i\frac{(x-x')^2}{4g(t-t')}
\right]
}{greefunczero}
Let us note that in principle the right hand side of
Eq.~\ceq{xitipsifin} should be multiplied by the determinant of the
operator $A^{-1}$, where
\beq{
A=i\frac{\partial}{\partial t}+g(\nabla+ib\nabla\phi)^2
}{operator}
However, it will be proved in the Appendix that $\det(A^{-1})=1$. The
reason, as 
explained in \cite{gi5}, is that in non-relativistic theories like
those treated here, there are no antiparticles and therefore charged
loops vanish identically. An explicit verification  that indeed
$\det(A^{-1})$ is trivial in theories in which the propagator
is proportional to $\theta(t-t')$ can be performed
following the procedure of Ref.~\cite{FePa}. 

Substituting in Eq.~\ceq{xitipsifin} the expressions of the currents
$J^*,J$ given in Eq.~\ceq{curcho}, we obtain:
\beq{
\Xi_T[\phi]=\exp\left\{
-\int dbd^dx\tilde{V}(b)e^{-ib({\phi(x)-\phi(x_0)})}e^{i\frac{(x-x_0)^2}{4gT}}
\right\}
}{xitiproved}
In the limit $T\longrightarrow +\infty$ the generating functional
$\Xi_T[\phi]$ of the fields $\psi^*,\psi$ together with the special
choice of currents \ceq{curcho} coincides exactly with the left hand
side of Eq.~\ceq{idefun}.

In conclusion, it has been shown that the partition function of the
nonlinear and nonlocal scalar field theory given in Eqs.~\ceq{partfun}
and \ceq{action} can be rewritten in the form of the equilibrium limit
of a local field theory:
\beq{
Z=\lim_{T\longrightarrow+\infty}\int{\cal D}\phi e^{-\int d^dx\left(
\frac 12(\nabla \phi)^2+\frac{m^2}{2}\phi^2
\right)}\Xi_T[\phi]
}{equlim}

\section{Non-perturbative approach}\label{sec2}
Let us write Eq.~\ceq{equlim} explicitly:
\begin{eqnarray}
Z&=&\lim_{T\to+\infty}\int{\cal D}\phi{\cal D}\psi^*{\cal
  D}\psi\exp\left\{
-\int d^dx\left(
\frac 12(\nabla\phi)^2+\frac {m^2}2\phi^2
\right)
\right\} \nonumber\\
&\times&\exp\left\{-i
\int dbdtd^dx\left[
i\psi^*\frac{\partial\psi}{\partial
  t}-g|(\nabla+ib\nabla\phi)\psi|^2-J^*\psi-J\psi^*
\right]
\right\}
\label{equlimfull}
\end{eqnarray}
In order to eliminate the field $\phi$, we introduce following
\cite{Femultifield} 
the
complex vector fields $\bm\xi^*,\bm\xi$ and express $Z$ as follows:
\begin{eqnarray}
Z&=&\lim_{T\longrightarrow+\infty}\int{\cal D}\phi{\cal D}\psi^*{\cal
  D}\psi {\cal D}\bm\xi^*{\cal D}\bm\xi\exp\left\{
-\int d^dx\left[
\frac 12(\nabla\phi)^2+\frac {m^2}2\phi^2
\right]
\right\}\nonumber\\
&\times&\exp\left\{-i
\int db dt d^dx\left[
i\psi^*\frac{\partial\psi}{\partial
  t}+g\bm\xi^*\cdot\bm\xi-g\bm\xi^*\cdot(\nabla+ib\nabla\phi) \psi
\right.
\right.\nonumber\\
&&\left.\left.\phantom{\frac{\partial\psi}{\partial
  t}} -g 
(\nabla-ib\nabla\phi) \psi^*\cdot\bm\xi-J^*\psi-J\psi^*
\right]\right\}\label{equlimfullb}
\end{eqnarray}
It is easy to check that Eq.~\ceq{equlimfull} is recovered after
eliminating the fields $\bm\xi^*,\bm\xi$ from Eq.~\ceq{equlimfullb}.
At this point we isolate in the expression of $Z$ the contribution due
to the field $\phi$:
\begin{eqnarray}
Z&=&\lim_{T\to+\infty}\int{\cal D}\psi^*{\cal D}\psi {\cal
  D}\bm\xi^*{\cal D}\bm\xi\exp\left\{
-i\int dbdtd^dx\left[
i\psi^*\frac{\partial\psi}{\partial
  t}+\right.\right.\nonumber\\
&+&\left.\left.\!\!\!\!\!\!\!\!\!\!\!\!\!\!
\phantom{*\frac{\partial\psi}{\partial
  t}}
g\bm\xi^*\cdot\bm\xi-g\bm\xi^*\cdot
\nabla\psi-g\bm\xi\cdot\nabla\psi^*-\psi J^*-\psi^*J
\right]
\right\}Z_\phi
\label{zphiinit}
\end{eqnarray}
where
\beq{
Z_\phi=\int{\cal D}\phi e^{-\int d^dx\left[
\frac 12(\nabla\phi)^2+\frac {m^2}{2}\phi^2-g\phi\nabla\cdot\int dbdt
b\left(
\bm\xi^*\psi-\bm\xi\psi^*
\right)
\right]}
}{zphicore}
$Z_\phi$ is the partition function of a free scalar field theory in
the presence of the external current:
\beq{
K(x)=\nabla\cdot\int dbdt b\left(
\bm\xi^*\psi-\bm\xi\psi^*
\right)
}{extcurr}
Let us note that this current is purely imaginary.
After performing the simple gaussian integration over $\phi$ we
obtain:
\beq{
Z_\phi=\exp\left\{
\frac{g^2}{2}\int d^d xd^dx'G(x,x')K(x)K(x')
\right\}
}{zphifinal}
$G(x,x')$ denotes the scalar field propagator:
\beq{
G(x,x')=\int\frac{d^dp}{(2\pi)^d} \frac {e^{-ip\cdot(x-x')}}{p^2+m^2}
}{greenfunctionscalarfields}
The total sign of the exponent appearing in the right hand side of
Eq.~\ceq{zphifinal} is negative. To show that, we put
$G(x,x')=\sum_{n=0}^{+\infty}\frac{\phi_n(x)\phi_n(x')}{\lambda_n}$,
where the $\phi_n(x)$'s are the eigenfunctions of the $d-$dimensional
differential operator $\Delta-m^2$ and the $\lambda_n$'s are their
respective eigenvalues. 
Thus Eq.~\ceq{zphifinal} may be rewritten as follows:
\beq{
Z_\phi=\exp\left\{
\frac{g^2}{2}\sum_{n=0}^{+\infty}\frac 1{\lambda_n}\left[\int d^d
  x\phi_n(x)K(x) \right]^2
\right\}
}{zphifinal2}
Due to the fact that the eigenvalues $\lambda_n$ are positive and
$K(x)$ is purely imaginary, the total sign of the exponent in the
above equation is negative as desired.

At this point it will be convenient to introduce the following
shorthand notation:
$\eta=x,b,t$ and $d^{d+2}\eta=d^dxdbdt$, so that
\beq{
Z_\phi=e^{\left\{
\frac{g^2}2\int d^{d+2}\eta d^{d+2}\eta'\big[
bb'G_{\mu\nu}(x,x')\big
(\xi^{*\mu}(\eta)\psi(\eta)-\xi^\mu(\eta)\psi^*(\eta) 
\big)
\big(\xi^{*\nu}(\eta')\psi(\eta')-\xi^\nu(\eta')\psi^*(\eta')\big)
\big]
\right\}}
}{zphifinal3}
where
\beq{
G_{\mu\nu}(x,x')=-\int\frac{d^dp}{(2\pi)^d} {e^{-ip\cdot(x-x')}}\frac{p_\mu
  p_\nu}{p^2+m^2} \qquad\qquad\mu,\nu=1,\ldots,d
}{gf2}
In writing Eq.~\ceq{zphifinal} we have used the explicit form of
the current $K(x)$ given in Eq.~\ceq{extcurr} and some integrations by
parts.
Substituting the expression of $Z_\phi$ of Eq.~\ceq{zphifinal3} back
in the original Eq.~\ceq{zphiinit}, the total partition function $Z$
becomes:
\beq{
Z=\lim_{T\to+\infty}\int{\cal D}\psi^*{\cal D}\psi{\cal
  D}\bm\xi^*{\cal D}\bm\xi 
e^{-iS_0+S_{int}}
}{zbeforeshift}
where
\beq{
S_0=\int d^{d+2}\eta\left[
i\psi^*\frac{
\partial\psi
}{\partial
  t}+g\bm\xi^*\cdot\bm\xi-g\bm\xi^*\cdot\nabla\psi-g\bm\xi\cdot\nabla
\psi^* -\psi J^*-\psi^*J
\right]
}{szerofree}
is the free part of the action, while the interaction term is:
\beq{
S_{int}=
\frac{g^2}2\int d^{d+2}\eta d^{d+2}\eta'\big[
bb'G_{\mu\nu}(x,x')\big
(\xi^{*\mu}(\eta)\psi(\eta)-\xi^\mu(\eta)\psi^*(\eta) 
\big)
\big(\xi^{*\nu}(\eta')\psi(\eta')-\xi^\nu(\eta')\psi^*(\eta')\big)
\big]
}{sint}
This is
the effective interaction resulting from the integration over the
fields $\phi$. Indeed, it is easy to realize that $S_{int}$ coincides
with the exponent of $Z_\phi$ in Eq.~\ceq{zphifinal3}.
The fact that this exponent is always negative assures the convergence
of the further integrations over the remaining fields.

In the free action $S_0$ of Eq.~\ceq{szerofree} the fields
$\bm\xi^*,\bm\xi$ and $\psi^*,\psi$ are coupled together. To
disentangle this unwanted coupling, we perform the following shift of
variables:
\beq{
\bm\xi^*=\nabla\psi^*+\delta\bm\xi^*
\qquad\qquad \bm\xi=\nabla\psi+\delta\bm\xi
}{shiftvar}
The new fields $\delta\bm\xi^*,\delta\bm\xi$ may be interpreted as the
fluctuations of the fields $\bm\xi^*,\bm\xi$ around their classical
configurations that are respectively given by $\nabla\psi^*$ and
$\nabla\psi$. 

Applying the shift \ceq{shiftvar} to Eq.~\ceq{zbeforeshift} we obtain:
\beq{Z=\lim_{T\to+\infty}\int{\cal D}\psi^*\int{\cal D}\psi\int{\cal D}(\delta\bm\xi^*)
\int{\cal D}(\delta\bm\xi)
e^{-iS_{0,\delta}+S_{int,\delta}}
}{sfhsdjfh}
Now the free action $S_{0,\delta}$ does not contain unwanted
interactions between the fields $\psi^*,\psi$ and the new fields
$\delta\bm\xi^*,\delta\bm\xi$:
\beq{
S_{0,\delta}=\int d^{d+2}\eta\left[
i\psi^*\frac{\partial\psi}{\partial
  t}-g\nabla\psi^*\cdot\nabla\psi+g\delta\bm\xi^*\cdot\delta\bm\xi-\psi
J^*-\psi^*J 
\right]
}{szerodelta}
The nonlinear part $S_{int,\delta}$ is given by:
\begin{eqnarray}
S_{int,\delta}&=&\frac{g^2}{2}\int d^{d+2}\eta d^{d+2}\eta'\Big[
bb'G_{\mu\nu}(x,x')\Big(
\partial^\mu\psi^*(\eta)\partial^\nu\psi^*(\eta')
\psi(\eta)\psi(\eta') +
\nonumber\\
\left.\right.
&+&\partial^\mu\psi(\eta)\partial^\nu\psi(\eta')
\psi^*(\eta)\psi^*(\eta') -2
\partial^\mu\psi^*(\eta)\partial^\nu\psi(\eta')
\psi(\eta)\psi^*(\eta')+\nonumber\\
&+&2\partial^\mu\psi^*(\eta)\delta\xi^{*\nu}(\eta')
\psi(\eta)\psi(\eta')
- 2\partial^\mu\psi^*(\eta)\delta\xi^{\nu}(\eta')
\psi(\eta)\psi^*(\eta')- \nonumber\\
&-&\delta\xi^{*\mu}(\eta')\partial^\nu\psi(\eta)
\psi(\eta)\psi^*(\eta')+
2\partial^\mu\psi(\eta)\delta\xi^{\nu}(\eta')
\psi^*(\eta)\psi^*(\eta')+\nonumber\\
&+&\delta\xi^{*\mu}(\eta)\delta\xi^{\nu}(\eta')
\psi(\eta)\psi(\eta')
+\delta\xi^{\mu}(\eta)\delta\xi^{\nu}(\eta')
\psi^*(\eta)\psi^*(\eta')-\nonumber\\
&-&2\delta\xi^{*\mu}(\eta)\delta\xi^{\nu}(\eta')
\psi(\eta)\psi^*(\eta')\Big)\Big]\label{fsdfsdfsdf}
\end{eqnarray}
where $\partial^\mu=\frac{\partial}{\partial x_\mu}$.
At this point we can expand the partition function $Z$ in powers of
the currents $J$ and $J^*$:
\begin{eqnarray}
Z&=&\lim_{T\to+\infty}\sum_{n=1}^\infty\sum_{m=1}^\infty\int d^{d+2}\eta_1
\cdots \int d^{d+2}\eta_n\int d^{d+2}\eta_1'\cdots\int
d^{d+2}\eta_m\times \nonumber\\
&\times& Z^{(nm)}(\eta_1,\ldots,\eta_n;\eta_1',\ldots,\eta_m')
J(\eta_1)\cdots J(\eta_n)J^*(\eta_1)\cdots J^*(\eta_n)
\label{expansion}
\end{eqnarray}
It is easy to check that in the above series many terms disappear in
the limit $T\longrightarrow+\infty$. 
They vanish due to the effect of the propagators of
the fields $\psi^*,\psi$ that are
given by
Eq.~\ceq{greefunczero}. In Eq.~\ceq{expansion}, due to the special
form of the currents $J^*,J$ defined in Eq.~\ceq{curcho}, these
propagators have to be evaluated in the special case $t-t'=T$.
As a consequence, each contraction of the fields $\psi^*,\psi$
generates a factor $T^{-\frac d2}$ and, for this reason,
many Feynman diagrams are suppressed in the limit
 $T\longrightarrow+\infty$.
Only those terms in which the factors
$T^{-\frac d2}$ are exactly compensated
by the positive powers of $T$ contained in  the currents
$J^*$ survive. 

It is also possible to show that the partition function
\ceq{sfhsdjfh} is independent of the value of the coupling constant
appearing in the actions $S_{0,\delta}$ and $S_{int,d}$ of
Eqs.~\ceq{szerodelta}--\ceq{fsdfsdfsdf} respectively. This could be
expected from the fact that the parameter $g$ does not appear in the
original model of Eqs.~\ceq{partfun}--\ceq{action}.
To prove that, we consider
the path integral
 $Z_g(T)$ in the left hand side of Eq.~\ceq{sfhsdjfh} before
taking the limit $T\longrightarrow+\infty$. 
Clearly $Z=\lim_{T\to+\infty}Z_g(T)=Z_g(+\infty)$.
Analogously, we will use the symbols
$S_{0,\delta}(g,T)$ and $S_{int,\delta}(g)$ for the actions
$S_{0,\delta}$ and $S_{int,\delta}$ in order to emphasize their
dependence on the parameters $g$ and $T$.
It will be shown in the following that
\beq{
Z_g(+\infty)=Z_{g'}(+\infty)
}{star}
even if $g$ and $g'$ do not coincide.
To begin with, we perform in the free action
$S_{0,\delta}(g,T)$ and in the interation part $S_{int,\delta}(g)$ the
time rescaling:
\beq{t=\frac{g'}{g}t'}{timerescaling}
After the above rescaling, $S_{0,\delta}(g,T)$ and $S_{int,\delta}(g)$ read
as follows:
\beq{
S_{0,\delta}(g',\frac{g}{g'}T)=\int dt'db d^dx\left[
i\psi^*\frac{\partial\psi}{\partial t}-g'\nabla\psi^*\cdot\nabla\psi
+g'\delta \bm \xi^*\cdot\delta\bm \xi-\frac{g'}{g}\psi J^*-\frac{g'}{g}
\psi^*J
\right]
}{szeroafterrescaling}
and
\begin{eqnarray}
S_{int,\delta}(g')&=&\frac{{g'}^2}{2}\int d^{d}xdbdt d^{d}x'db'dt'\Big[
bb'G_{\mu\nu}(x,x')\Big(
\partial^\mu\psi^*(\eta)\partial^\nu\psi^*(\eta')
\psi(\eta)\psi(\eta') +
\nonumber\\
\left.\right.
&+&\partial^\mu\psi(\eta)\partial^\nu\psi(\eta')
\psi^*(\eta)\psi^*(\eta') -2
\partial^\mu\psi^*(\eta)\partial^\nu\psi(\eta')
\psi(\eta)\psi^*(\eta')+\nonumber\\
&+&2\partial^\mu\psi^*(\eta)\delta\xi^{*\nu}(\eta')
\psi(\eta)\psi(\eta')
- 2\partial^\mu\psi^*(\eta)\delta\xi^{\nu}(\eta')
\psi(\eta)\psi^*(\eta')- \nonumber\\
&-&\delta\xi^{*\mu}(\eta')\partial^\nu\psi(\eta)
\psi(\eta)\psi^*(\eta')+
2\partial^\mu\psi(\eta)\delta\xi^{\nu}(\eta')
\psi^*(\eta)\psi^*(\eta')+\nonumber\\
&+&\delta\xi^{*\mu}(\eta)\delta\xi^{\nu}(\eta')
\psi(\eta)\psi(\eta')
+\delta\xi^{\mu}(\eta)\delta\xi^{\nu}(\eta')
\psi^*(\eta)\psi^*(\eta')-\nonumber\\
&-&2\delta\xi^{*\mu}(\eta)\delta\xi^{\nu}(\eta')
\psi(\eta)\psi^*(\eta')\Big)\Big]\label{sintdeltaafterrescaling}
\end{eqnarray}
We remark that in the above equation 
$t$ and $t'$ denote
the new time variable
obtained after the rescaling of Eq.~\ceq{timerescaling}.
Moreover,
the fields depend on the time $t'$ multiplied by the scaling factor
$\frac {g'}{g}$,
i.~e. $\eta=\frac {g'}{g}t,x,b$ and
$\eta'=\frac {g'}{g}t',x,b$.
Apart from this implicit dependence, the parameter $g$ appears
also explicitly in the current term of the free action of
Eq.~\ceq{szeroafterrescaling}.
There is no other dependence on $g$ both in the free action and in the
interaction term of Eq.~\ceq{sintdeltaafterrescaling}.

It turns out that 
the presence of $g$ in the current term is limited to a
factor $\frac{g}{g'}$ which rescales the time $T$.
To show that, we write down the expression of this current term,
which, apart
from an irrelevant overall constant, is
equal to:
\begin{eqnarray}
S_{curr,\delta}(g',\frac{g}{g'}T)&=&\int db dt'd^dx\frac{g'}{g}\left[
\left(
4\pi ig'\left(
\frac{g}{g'}T
\right)
\right)^{\frac d2}\delta\left(
\frac{g'}{g}t'
\right)\psi(\frac{g'}{g}t',x,b)+\right.\nonumber\\
&+&\left.\delta^{(d)}(x-x_0){\tilde
  V}(b)\delta\left(
\frac{g'}{g}t'-T
\right) \psi^*(\frac{g'}{g}t,x,b)
\right]
\label{scurrdelta}
\end{eqnarray}
Using the following identities between dirac delta functions
\beq{
\delta(\frac{g'}{g}t')=\frac{g}{g'}\delta(t')\qquad\qquad
\delta\left(\frac{g'}{g}t'-T \right)=\frac{g}{g'}\delta\left(
t'-\frac{g}{g'}T
\right)
}{deltafunctionsidentities}
the expression of $S_{curr}(g',\frac{g}{g'}T)$ becomes:
\begin{eqnarray}
S_{curr,\delta}(g',\frac{g}{g'}T)&=&\int db dt'd^dx\left[
\left(
4\pi ig'\left(
\frac{g}{g'}T
\right)
\right)^{\frac d2}\delta(t')\psi(\frac{g'}{g}t',x,b)+\right.\nonumber\\
&+&\left.\delta^{(d)}(x-x_0){\tilde
  V}(b)\delta\left(
t'-\frac{g}{g'}T
\right) \psi^*(\frac{g'}{g}t,x,b)
\right]
\label{scurrdelta2}
\end{eqnarray}
It is clear from the above equation that, as predicted,
 the old coupling constant $g$ enters in
current term $S_{curr,\delta}(g',\frac{g}{g'}T)$ 
only inside the scaling factor
$\frac{g}{g'}$ of the time $T$. 
There is no other explicit dependence 
on $g$ in the action. In fact, if we put $J=J^*=0$ 
it is easy to realize that
the old coupling constant $g$ has been already replaced by
$g'$ in
 the free action 
of
Eq.~\ceq{szeroafterrescaling} and in the interaction term
of Eq.~\ceq{sintdeltaafterrescaling}.

Of course we have to remember that, after the time rescaling
of Eq.~\ceq{timerescaling},
$g$ is still appearing inside the fields, because
their dependence on the time variable is of the form
$\psi^*=\psi^*(\frac{g'}{g}t',x,b)$ and
$\psi=\psi(\frac{g'}{g}t',x,b)$. Analogous equations are valid for
$\delta\bm \xi^*$ and $\delta\bm \xi$. However, since we have to
perform a path integration over all field configurations, this
implicit
 presence
of $g$ may be easily eliminated inside the path integral
by the change of variables:
\begin{eqnarray}
\psi^{*\prime}(t',x,b)=\psi(\frac{g'}{g}t',x,b)&\qquad\qquad&
\psi^{*\prime}(t',x,b)=\psi(\frac{g'}{g}t',x,b)\\
\bm\xi^{*\prime}(t',x,b)=\bm\xi(\frac{g'}{g}t',x,b)&\qquad\qquad&
\bm\xi^{*\prime}(t',x,b)=\bm\xi(\frac{g'}{g}t',x,b)
\end{eqnarray}

Summarizing, 
we are able to write the following identity:
\beq{
Z_g(T)=Z_{g'}(\frac{g}{g'}T)
}{doppioquadrato}
where
\beq{Z_{g'}(\frac{g}{g'}T)=
\int{\cal D}\psi^{*\prime}
\int{\cal D}\psi'\int{\cal D}(\delta\bm\xi^{*\prime})
\int{\cal D}(\delta\bm\xi')
e^{-iS_{0,\delta}(g',\frac{g}{g'}T)+S_{int,\delta}(g')}
}{doppiastar}
The actions $S_{0,\delta}(g',\frac{g}{g'}T)$ and
$S_{int,\delta}(g')$ in Eq.~\ceq{doppiastar} are given by:
\begin{eqnarray}
S_{0,\delta}(g',\frac{g}{g'}T)&=&
\int dt'dbd^{d}x\left[
i\psi^{*\prime}\frac{\partial\psi'}{\partial
  t'}-g'\nabla\psi^{*\prime}
\cdot\nabla\psi'+g'\delta\bm\xi^{*\prime}\cdot\delta\bm\xi'-
\right.\nonumber\\
\!\!\!\!\!\!\!\!\!\!\!\!
&\!\!\!\!\!\!\!\!\!\!\!\!\!\!\!\!\!\!\!\!-&
\!\!\!\!\!\!\!\!\!\!\!\!
\left.\left(
4\pi i g'(\frac{g}{g'}T)
\right)^{\frac d2}\delta(t')\psi'(t',x,b)
+\delta^{(d)}(x-x_0){\tilde V}(b)\delta(t'-\frac{g}{g'}T)
\psi^{*\prime}(t',x,b)
\right]\label{triploquadrato}
\end{eqnarray}
and
\begin{eqnarray}
S_{int,\delta}(g')&=&\frac{{g'}^2}{2}\int d^{d}xdbdt d^{d}x'db'dt'\Big[
bb'G_{\mu\nu}(x,x')\Big(
\partial^\mu\psi^{*\prime}(\eta)\partial^\nu\psi^{*\prime}(\eta')
\psi'(\eta)\psi'(\eta') +
\nonumber\\
\left.\right.
&+&\partial^\mu\psi'(\eta)\partial^\nu\psi'(\eta')
\psi^{*\prime}(\eta)\psi^{*\prime}(\eta') -2
\partial^\mu\psi^{*\prime}(\eta)\partial^\nu\psi'(\eta')
\psi'(\eta)\psi^{*\prime}(\eta')+\nonumber\\
&+&2\partial^\mu\psi^{*\prime}(\eta)\delta\xi^{*\prime\nu}(\eta')
\psi'(\eta)\psi'(\eta')
- 2\partial^\mu\psi^{*\prime}(\eta)\delta\xi^{\prime\nu}(\eta')
\psi'(\eta)\psi^{*\prime}(\eta')- \nonumber\\
&-&\delta\xi^{*\prime\mu}(\eta')\partial^{\prime\nu}\psi'(\eta)
\psi'(\eta)\psi^{*\prime}(\eta')+
2\partial^\mu\psi'(\eta)\delta\xi^{\prime\nu}(\eta')
\psi^{*\prime}(\eta)\psi^{*\prime}(\eta')+\nonumber\\
&+&\delta\xi^{*\prime\mu}(\eta)\delta\xi^{\prime\nu}(\eta')
\psi'(\eta)\psi'(\eta')
+\delta\xi^{\prime\mu}(\eta)\delta\xi^{\prime\nu}(\eta')
\psi^{*\prime}(\eta)\psi^{*\prime}(\eta')-\nonumber\\
&-&2\delta\xi^{*\prime\mu}(\eta)\delta\xi^{\prime\nu}(\eta')
\psi'(\eta)\psi^{*\prime}(\eta')\Big)\Big]\label{triplocerchio}
\end{eqnarray}
where now $\eta=(t,x,b)$ and $\eta'=(t',x,b)$.
As we see from Eqs.~\ceq{doppioquadrato} and
\ceq{triploquadrato}--\ceq{triplocerchio}, the only left dependence on
$g$ is in the rescaled time $\frac {g}{g'}T$ contained
in the current term. In
the limit 
$T\longrightarrow +\infty$, of course, $\frac{g}{g'}\infty=\infty$,
i. e.:
\beq{
Z_g(+\infty)=Z_{g'}(+\infty)
}{finalresssss}
Since by definition $Z_g(+\infty)=Z$, where $Z$ is the partition
function of Eq.~\ceq{sfhsdjfh}, we have shown that 
$Z$ does not depend on the value of the parameter $g$. This concludes
our proof.

\section{Mean field approximation}
In order to proceed, we treat the $\delta\bm\xi^*,\delta\bm\xi$ fields
in a mean 
field theory approximation, i.~e. assuming that the density of these
fields exhibits only little deviations from the average value.
Exploiting the fact that only the correlator $\langle
\delta\xi_\mu^*(\eta)\delta\xi_\nu(\eta')\rangle=
\frac{\delta_{\mu\nu}}{g}\delta(\eta-\eta')$ is different from zero, 
where $\delta(\eta-\eta')=\delta(x-x')\delta(b-b')\delta(t-t')$, it is
easy to check that the mean field effective action is given by:
\beq{
S_\delta^{MF}=S_{0,\delta}^{MF}+S_{int,\delta}^{MF}
}{effactmft}
with
\beq{
S_{0,\delta}^{MF}=\int d^{d+2}\eta\left[
i\psi^*\frac{\partial\psi}{\partial
  t}-g\nabla\psi^*\cdot\nabla\psi-\psi
J^*-\psi^*J
\right]
}{s0deltamft}
and
\begin{eqnarray}
S_{int,\delta}^{MF}&=&\frac{g^2}{2}\int d^{d+2}\eta d^{d+2}\eta'
bb'G_{\mu\nu}(x,x')\left[\phantom{\frac{2}g}\!\!\!\!\!
\left(
\partial^\mu\psi^*(\eta)\psi(\eta)-\partial^\mu\psi(\eta)\psi^*(\eta)
\right)
\right.\nonumber\\
&&\left.\left(
\partial^\nu\psi^*(\eta')\psi(\eta')-\partial^\nu\psi(\eta')\psi^*(\eta')
\right)-\frac{2}g\delta^{\mu\nu}\delta(\eta-\eta')\psi^*(\eta')\psi(\eta)
\right]\label{cutreg}
\end{eqnarray}
The presence of the Dirac delta function in the last term requires the
computation of the propagator $G_{\mu\nu}(x,x')$ at coinciding points 
$x=x'$. Since $G_{\mu\nu}(x,x)$ is divergent when $d>1$, we regularize
this singularity using the dimensional regularization. After a few
computations one finds:
\beq{
G_{\mu\nu}(x,x)=\frac{\delta_{\mu\nu}}{(4\pi)^{\frac d2}}m^d\frac
d2\Gamma(-\frac d2)
}{counterterm}
where $\Gamma(z)$ is the gamma function.
In the case $m=0$ in which the scalar field $\phi$ becomes massless,
$G_{\mu\nu}(x,x)$ vanishes identically, so that the introduction of
counterterms is not necessary. For $m\ne 0$ and odd dimensions, the
right hand side of Eq.~\ceq{counterterm} does not vanish, but it is
regular and once again no counterterms are needed.
Singularities appear only when the scalar field is massive and the
number of dimensions is odd. For instance, if $d=2$, we obtain from
Eq.~\ceq{counterterm}:
\beq{
G_{\mu\nu}(x,x)=-\frac{
m^2}\epsilon\frac{\delta_{\mu\nu}}{2\pi}
+\mbox{finite}
\qquad\qquad\epsilon=d-2
}{ccdfsdf}
This singularity gives rise in the action $S_{int,\delta}^{MF}$
 to the term $\frac {gm^2}{2\pi\epsilon}\int d^{d+2}\eta b^2
 \psi^*(\eta)\psi(\eta)$ that can be reabsorbed by adding a suitable
 mass conterterm for the fields $\psi^*,\psi$ in the free action
 $S_{0,\delta}^{MF} $.
\section{Conclusions}
In this letter we have considered a class of massive scalar field
theories with  potentials of the kind given in Eq.~\ceq{vecpotfouref}.
It has been shown that these theories, which
appear to be
intractable with the usual techniques, see for instance 
the potential in Eq.~\ceq{exaone}, can be
casted in a form that resembles that of a standard
field theory. 
Indeed, the partition function defined in
Eqs.~(\ref{sfhsdjfh}--\ref{fsdfsdfsdf}) is that of an usual complex
scalar field theory coupled to the vector fields
$\delta\bm\xi^*,\delta\bm\xi$. 
We have treated  here only the partition function of the scalar fields
$\phi$, but it is not
difficult to extend our result also to their generating 
functional. The only difference is that in this case
one should add to the current $K(x)$ of Eq.~\ceq{extcurr} also the
external current of the fields $\phi$.

Of course, given the complexity of the
original field  theories discussed here, one cannot expect that they
become exactly solvable after the mapping explained in Section II.
However, if the potential $V(\phi(x)-\phi(x_0))$ is small, for
instance because it is multiplied by an overall small constant
$k$, then 
perturbation theory may be attempted. In fact, the Fourier transform
$\tilde{V}(b)$
of that potential, which is small too, is only present in the current
$J(\eta)$, see 
Eq.~\ceq{curcho}.  For this reason, the 
terms of $n-$th order in the perturbative
expansion in the coupling constant $k$ simply coincides with the power 
$n$ of the currents
$J(\eta_1)\cdots J(\eta_n)$ in the series of Eq.~\ceq{expansion}.
This perturbative strategy was clearly not possible
in the starting partition function of the scalar fields of
Eq.~\ceq{partfun}, because in the potential $V(\phi(x)-\phi(x_0))$
contains in the most general case an infinite number of powers of
$\phi$, see the example of Eq.~\ceq{exaone}. In a similar way, for
large values of the coupling constant, one may use a strong coupling
expansion, as explained for instance in Refs.~\cite{kleinert}.
One should instead resist
the temptation of using as a perturbative parameter the fictitious
coupling constant
$g$ which is present in the action $S_{int,\delta}$.
As mentioned in the previous Section, in fact, this parameter
disappears from the theory after performing the limit
$T\longrightarrow+\infty$. This is understandable, because $g$ was not
 the original scalar field theory.

We have also explored a nonperturbative approach, in which the
original scalar fields are integrated out exactly and a mean field
approach is applied to the resulting
vector fields $\delta\bm\xi^*,\delta\bm\xi$.
In doing that, one finds that the theory
is affected by singularities which are cured using the dimensional
regularization. The form of the counterterm which is necessary to
absorb these singularities has been computed.
Let us note that this counterterm does not exhaust all the
renormalizability issues of the theory expressed by the action
$S_{\delta}^{MF}$  given by Eqs.~(\ref{effactmft}--\ref{cutreg}).
First of all, there are still divergences that may come from the
interactions of the fields $\psi^*,\psi$. Moreover, there are also the
singularities connected with the presence of the potential
$V(\phi(x)-\phi(x_0))$. These divergences are apparently hidden due to
the fact that, thanks to the methods of Section II, it has been
possible to confine to the current $J(t,x,b)$ of Eq.~\ceq{curcho} all
the dependencies on the potential. This does not mean however that
this interaction has now become harmless. To convince oneself that
this is not the case, it is
sufficient to give a glance to at the expansion of the partition
function $Z$ given in Eq.~\ceq{expansion}. There, the convergence of
the integrations
over the variables $b_1,\ldots,b_n$ strongly depends on the form of
the potential $\tilde{V}(b)$ that appears inside the currents
$J(\eta_1),\ldots,J(\eta_n)$.
It is impossible to formulate a renormalization theory like that of
usual field theories with polynomial interactions in the case of
general $d-$dimensional models such as those discussed in this work.
For this reason, in the future it will be necessary to identify
particular examples of potentials that are physically relevant and to
investigate renormalization issues in those special cases.
\begin{appendix}
\section{Proof of the triviality of the determinant of the operator
  $A^{-1}$ of Eq.~\ceq{operator}}
In this Appendix we shot that the inverse determinant
\beq{
{\det}^{-1}\left[
i\frac{\partial}{\partial t}+(\nabla+ib\nabla\phi)^2)
\right]=\int{\cal D}\psi^*{\cal D}\psi\exp\left\{
-i\int dbdtd^dx\left[
i\psi^*\frac{\partial\psi}{\partial t}-g|(\nabla+ib\nabla\phi)\psi|^2
\right]
\right\}
}{detdef}
is trivial. To this purpose, let us plit the action appearing in the
exponent of the right hand side of Eq.~\ceq{detdef} into a free and an
interaction part:
\beq{
S=-i\int dbdtd^dx\left[
i\psi^*\frac{\partial \psi}{\partial t}-g|(\nabla+ib\nabla\phi)\psi|^2
\right]=S_0+S_I
}{splitting}
where:
\beq{
S_0=\int dbdtd^dx\left[
i\psi^*\frac{\partial \psi}{\partial t}-g\nabla\psi^*\nabla\psi
\right]
}{szero}
and
\beq{
S_I=\int db dt d^dx\left[
igb\nabla\phi(\psi^*\nabla\psi-\psi\nabla\psi^*)-gb^2(\nabla\phi)^2|\psi|^2
\right]
}{sinteraction}
At the tree level the relevant Feynman diagrams of this theory are
shown in Figs.~1 and 2a-2c.
\begin{figure}
\includegraphics[height=.5cm]{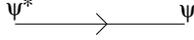}\\
\caption{Propagator of the fields $\psi^*,\psi$.}
\label{fig1}
\end{figure}
\begin{figure}
\includegraphics[height=3cm]{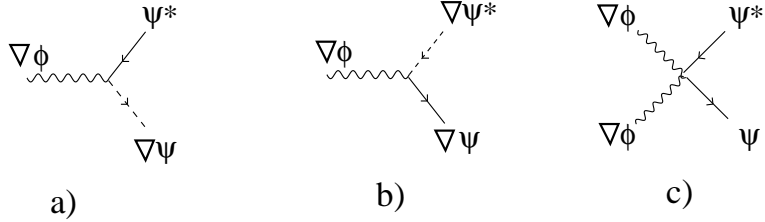}\\
\caption{Vertex diagrams corresponding to the action \ceq{sinteraction}.}
\label{fig2}
\end{figure}
The propagator of Fig.~\ref{fig1} is given by:
\beq{\langle\psi^*(t',x',b')\psi(t,x,b)\rangle=-\frac{i\theta(t-t')}{|4\pi
    i g(t-t')|^{\frac d2}}\exp\left[
\frac{i(x-x')^2}{4g(t-t')}
\right]\delta(b-b')
}{propagator}
We may now expand the right hand side of Eq.~\ceq{detdef} in powers of
$g$.
Apart from the zeroth order, all Feynman diagrams are closed one loop
diagrams in which the internal legs propagate the
fields $\psi^*,\psi$, while the external legs propagate the
field $\phi$.
At order $n$ with respect to $g$ these Feynman diagrams are generated
from the contraction of
pairs of the $\psi^*,\psi$ fields inside products of $n$ vertices
which, in their general form, look as follows:
\begin{eqnarray}
\Gamma_{I,n}&=&\int db_1 dt_1d^dx_1\cdots\int db_n dt_nd^dx_n\cdots
igb_i\partial_\mu\phi(x_i)\psi^*(t_i,x_i,b_i)\partial^\mu
\psi(t_i,x_i,b_i)\cdots\\\nonumber 
&&\cdots(-i)gb_j\partial_\nu\phi(x_j)\psi(t_j,x_j,b_j)\partial^\nu\psi^*(t_j,
x_j,b_j)\cdots(-g)b_k^2(\nabla\phi(x_k))^2|\psi(t_k,x_k,b_k)|^2\cdots
\label{genverdia}
\end{eqnarray}
Here the indices $i,j,k$ are such that $1\le i< j<k\le n$. 
The number $I$ of external
legs depends on the number of vertices of the type of
Fig.~\ref{fig2}-c which appear in  $\Gamma_{I,n}$ and
ranges within the interval:
\beq{
n\le I\le 2n
}{range}
A graphical representation of the connected diagrams which are
associated with
$\Gamma_{I,n}$ is given in
Fig.~\ref{fig3}.
\begin{figure}
\includegraphics[height=3cm]{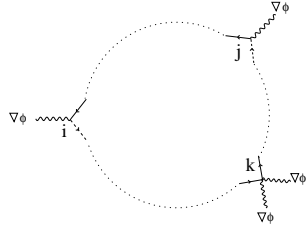}\\
\caption{Graphical representation of a general connected diagram 
coming out from the contraction of the
$\psi^*,\psi$ fields inside the product of vertices
  $\Gamma_{I,n}$ of Eq.~\ref{genverdia}.} 
\label{fig3}
\end{figure}
At this point we note that the pairs of fields $\psi^*,\psi$ in
$\Gamma_{I,n}$ may be contracted in a $(n-1)!$ number of ways.
Thus, $\Gamma_{I,n}$ consists in a sum of $(n-1)!$ Feynman diagrams. Let
$\Gamma_{I,n,\sigma}$ be one of these diagrams.
$\sigma$ denotes an arbitrary permutation acting on the set of $(n-1)$
indices 
$\{2,3,\ldots,n\}$. The expression of $\Gamma_{I,n,\sigma}$ may be
obtained by contracting the field $\psi^*$ with the field $\psi$ of
the $\sigma(2)-$th vertex. Next, the field $\psi^*$ of the
$\sigma(2)-$th vertex will be contracted with the field $\psi$ of the
$\sigma(3)-$th vertex and so on. $\sigma(i)$, $i=1,\ldots,n$ denotes
here the result 
of the permutation of the $i-$th index.
Since there are $(n-1)!$ permutations $\sigma$ of this kind, it is easy to
check that in this way it is possible to compute all the $(n-1)!$
contributions to $\Gamma_{I,n}$. Let's check now more in details the
structure of each diagram $\Gamma_{I,n,\sigma}$.
Due to the particular form of the propagator \ceq{propagator},
$\Gamma_{I,n,\sigma}$ will be proportional to the following product of
Heaviside $\theta-$functions:
$\theta(t_1-t_{\sigma(2)})\theta(t_{\sigma(2)}-t_{\sigma(3)})\cdots
\theta(t_{\sigma(n)}-t_1)$.
The above product of Heaviside $\theta-$functions enforces the
condition:
\beq{
t_1>t_{\sigma(2)}>t_{\sigma(3)}>\cdots>t_{\sigma(n)}>t_1
}{sequence}
Clearly, this sequence of inequalities is impossible. For this reason,
the products of Heaviside $\theta-$functions vanishes identically. 
As
a consequence, the determinant of Eq.~\ceq{detdef} is trivial, i. e.:
\beq{
\det\left[
i\frac{\partial}{\partial t}+(\nabla+ib\nabla\phi)^2)
\right]=1
}{dettriv}
 because
all its contributions vanish identically apart from the case
$n=0$.
This result could be expected from the fact that the field theory
given in Eq.~\ceq{detdef} is a particular case of a nonrelativistic
complex scalar field theory. It is indeed well known that
these nonrelativistic field theory give rise to nontrivial
determinants \cite{gi5}.
\end{appendix}

\end{document}